\documentclass[a4paper,12pt]{article}

\usepackage{amsmath}
\usepackage{braket}
\usepackage{amssymb}
\usepackage{amsfonts}
\usepackage{latexsym}
\usepackage{graphicx}
\usepackage{latexsym}
\usepackage{color}
\usepackage{indentfirst}
\usepackage{hyperref}
\usepackage{verbatim}
\usepackage{multicol}

\def\ln{\,\mbox{ln}\,}

\renewcommand{\vec}[1]{{\bf #1}}

\def\beq{\begin{eqnarray}}
\def\eeq{\end{eqnarray}}

\begin{document}

\begin{center}

{\Large\textbf{Scalar field theory for warm dark matter}}

\vspace{2cm}

\textbf{Jhonny A. A. Ruiz}$^{a,b}$

\vspace{1cm}

\small{\sl
(a) {N\'ucleo Cosmo-ufes \& PPGCosmo, Universidade Federal do Esp\'irito Santo,\\
29075-910, Vit\'oria, ES, Brazil}
\vskip 2mm

(b) {Departamento de F\'{\i}sica,  ICE,  Universidade Federal de Juiz de Fora,\\
36036-330 Juiz de Fora, \ MG, \ Brazil}
\vskip 2mm


\vskip 5mm

{\sl E-mail:
jaar@cosmo-ufes.org}
}

\end{center}


\vskip 6mm
\begin{quotation}
\noindent
{\large {\it Abstract}}.
\quad
The warm dark matter (WDM) can be described by a simple and useful model called reduced relativistic gas (RRG).  In this work, it is analytically constructed the scalar field actions minimally and non-minimally coupled to gravity, which are equivalent to RRG in the sense they produce the same cosmological solutions for the conformal factor of the metric. In particular, we construct the scalar theory which corresponds to the model of an ultra-relativistic ideal gas of spinless particles possessing conformal symmetry. Finally, the possibility of supplementing our scalar field model for WDM with dynamical dark energy in the form of a running cosmological constant (RCC) is also considered.

\end{quotation}
\vskip 4mm

\section{Introduction}\label{Sec1}
The standard model of cosmology ($\Lambda$CDM) describes the properties and evolution of the universe as a whole, considering the existence of a new kind of cosmic fluids known as dark energy (DE) and dark matter (DM) and whose theoretical predictions are in very good agreement with observational data \cite{Planck2018VI,anderson2014clustering}. However, although this model seems to pass most of the experimental tests, there are still some discrepancies and tensions with data that cannot be explained naturally within this framework \cite{buchert2016observational}.

In the $\Lambda$CDM model, the role of DE is assumed by the positive cosmological constant (CC) $\Lambda$, which is considered as a fluid with negative pressure, as the most natural and simple explanation for the current accelerating phase \cite{caldwell1998cosmological}-\cite{peebles2003cosmological}. Nonetheless, this leads to the well known CC problem, opening new searches and possibilities for the solution of the DE problem \cite{weinberg1989cosmological,sola2013cosmological}. These new developments have also given rise to some extensions of the standard model or even modifications of the gravitational theory which is based on \cite{nojiri2007introduction,capozziello2011extended}.

One of these natural extensions consists of considering the possible running of $\Lambda$, where this running emerges as a consequence of the quantum corrections in the effective action and its simplest version can be described through a minimal subtraction scheme in curved spacetime \cite{nelson1982scaling,bukhbinder1984renormalization}. At low energies (IR), it should be taken into account the decoupling of the massive fields, however, this decoupling cannot be verified for the $\Lambda$ case \cite{gorbar2003renormalization},
but the non-running can be proved neither \cite{shapiro2009possible}. Thus, one can explore the running CC only in the phenomenological setting. Despite this fundamental limitation, these running vacuum models (RVM) have shown interesting features for both background and perturbation levels, fitting quite well the observational data, and in some cases, with more accuracy than standard model \cite{grande2010cosmic}-\cite{basilakos2020gravitational}.

Another common possibility for explaining the current and primordial properties of an expanding universe, it can be given in terms of a scalar field, where this field can be minimally or non-minimally coupled to gravity, it has a time-dependent equation of state and its dynamical evolution can be compatible with the solution of the DE problem and even it can provide a suitable inflationary dynamics with excellent confidence level according to observations \cite{peebles1988cosmology}-\cite{copeland2006dynamics}.

Now, coming back to our initial discussion about the standard model, it is also well known that the dark matter (DM) component is still an open question. DM, considered within $\Lambda$CMD as an exotic and non-relativistic matter, with zero pressure, is necessary for a consistent explanation of large scale structure (LSS) formation and evolution in an expanding universe and also for describing baryon acoustic oscillations (BAO) and cosmic microwave background (CMB) anisotropies \cite{bertone2005particle}-\cite{aghanim2020planck}. Nowadays there are several candidates for CDM particle \cite{bergstrom2000non}, however, the assumption of this DM component as cold leads some inconsistencies with data \cite{warren2006precision}-\cite{boylan2011too}. Therefore, the DM component could be hot (HDM) or warm (WDM) instead of cold. HDM scenarios are today ruled out by new cosmic data-sets, thus letting the WDM as a more realistic alternative \cite{hannestad2000self}-\cite{viel2005constraining}.

Considering WDM usually implies the complete solution of Boltzmann equations, taking into account the possible WDM candidates and their specific properties, but this process is rather robust and it can be complicated from the analytical point of view. As a useful alternative, the reduced relativistic gas (RRG) model emerges as a very simple and model-independent method for describing WDM, where this model consists in a simplification of the general model for relativistic gas of massive particles, where it is considered the same value of kinetic energy for all of them \cite{juttner1911dynamik}\cite{sakharov1966initial}. The RRG model represents just one example of the possible cosmological alternatives equations of state, where their characteristic parameter $\omega$ is not constant but may have different forms  \cite{gorini2004tachyons}\cite{shapiro2002massive}. The main property of the RRG model in cosmology is WDM description by one single free parameter $b$, which characterizes its warmness, without considering interaction with other forms of matter \cite{fabris2009dm}\cite{fabris2012testing}\cite{hipolito2018general}.

We are interested here in reconstructing and discussing a consistent scalar field theory for describing WDM using the useful approximation of the simple RRG model, as well as its cosmological consequences. For this purpose, we consider the description in terms of a minimal and non-minimal scalar field using the hydro-dynamical approach and the properties of the conformal transformation. Also, the possible inclusion of a dynamical dark energy component in the form of an RCC is considered.

The paper is organized as follows. In the next sect. \ref{Sec2}, we summarize the main properties of the simple RRG model for warm dark matter.  In sect. \ref{Sec3}, we formulate the scalar field theory for this WDM for the case of the minimally coupled scalar field to gravity. Later, in sect. \ref{Sec4}, we extend this scalar field theory to the case of conformal coupling to gravity (non-minimal coupling), mapping both of the models using the properties of conformal transformation. Finally, in sect. \ref{Sec5} we explore the possibility of including a dynamical dark energy component using the QFT-inspired RCC model. We draw our conclusions and perspectives in sect. \ref{Sec6}.


\section{Standard cosmological model with RRG}\label{Sec2}
The cosmological models based on RRG were described in detail in \cite{de2005simple}. Let us give a brief account of the main relations which will be needed in what follows. The equation of state for the RRG model is given by
\begin{equation}\label{estadorrg}
p_R=\frac{\rho_R}{3}\left(1-s\right),\qquad\qquad s=\frac{\rho_{d}^{2}}{\rho_{R}^{2}}
\end{equation}
where $\rho_{d}\equiv mc^{2}N/V=\rho_{1}\left(1+z\right)^3$ is its rest energy density. Here, $m$, $c$ and $V$, are the mass of the particles, the light velocity and the volume associated to the gas, respectively. For the simple RRG cosmological model, we have the Friedmann equations
\begin{equation}\label{friedmannrrg}
\rho_R=\frac{3H^2}{\kappa^2},\qquad\qquad p_{R}=-\frac{1}{\kappa^2}\left(3H^{2}+2\dot{H}\right),
\end{equation}
for a spatially flat FRLW metric with line element
\begin{equation}\label{metric}
        ds^2=-dt^2+a^2(t)d\vec{x}^2,
\end{equation}
where $a(t)$ is the scale factor and the dot denotes derivatives with respect to the cosmic time $t$. We also have the correspondent continuity equation
\begin{equation}\label{RRGcosmeq1}
\frac{d\rho_R}{da}+\frac{(4-s)}{a}\rho_R=0,
\end{equation}
whose solution can be cast in the form
\begin{equation}\label{RRGcosmeq2}
\rho_{R}(a)=\frac{\rho_1}{a^3}\sqrt{1+\frac{b^2}{a^2}}
\end{equation}
where
\begin{equation}\label{21}
\rho_{1}=mc^{2}N/V_{0},\qquad\qquad b^2=\frac{\rho_{2}^{2}}{\rho_{1}^{2}}
\end{equation}
Here $\rho_1$ is the rest energy at some initial point $V=V_0$ and $\rho_{2}$ can be interpreted as the radiation component of the RRG \cite{de2005simple}. The warmness parameter $b$ can also be written in the form
\begin{equation}
b = \frac{v/c}{\sqrt{1-\frac{v^2}{c^2}}}
\end{equation}
such that, for a low warmness $\,v/c \ll 1$, we have $\,b \sim v/c $.
Thus, $\,b \approx  0$ means that the matter contents is ``cold''. The non-relativistic (NR) and ultra-relativistic (UR) limits correspond to the cases when $\rho_2=0$ and $\rho_1=0$, respectively, setting up the well known interpolation between radiation and dust components for the RRG \cite{fabris2009dm}. We can also write these limits in the form
\begin{equation}
UR:\begin{cases}
\rho_{1}\rightarrow 0\\
a\rightarrow 0
\end{cases}\Rightarrow \frac{a^2}{b^2}\ll 1,
\end{equation}
\begin{equation}\label{limits}
NR:\begin{cases}
\rho_{2}\rightarrow 0\\
a\rightarrow\infty
\end{cases}\Rightarrow \frac{b^2}{a^2}\ll 1
\end{equation}
such that, the equation of state parameter $\omega_{R}(\rho)=p_{R}/\rho_{R}$ for this RRG satisfies
\begin{equation}\label{omegar}
\lim_{a\to\infty}\omega_{R}(\rho)=\omega_{nr}(\rho)=0, \qquad \lim_{a\to0}\omega_{R}(\rho)=\omega_{ur}(\rho)=\frac{1}{3}
\end{equation}
representing exactly the aforementioned interpolation. Additionally, in this UR limit, the trace of the energy-momentum tensor would be
\begin{equation}
(T_{ur})^{\mu}_{\mu}=3p-\rho=\rho\left[3\omega_{ur}(\rho)-1\right]=0
\end{equation}
so we want to construct a scalar field theory for the RRG also satisfying these properties.\\ 
The RRG equation of state and other relations like continuity and warmness equations indicate that this is the model describing an ideal, this is, non-interacting gas of relativistic massive particles. Therefore, if thinking about its mapping to the theory of the scalar field, in the zero-order approximation our physical insight suggests that such a scalar model should correspond to a free scalar theory. On the other hand, since the RGG slightly deviates from the Juttner model \cite{juttner1911dynamik}, there can be a certain defect or deviation from the free scalar field. Thus, the conformal potential should take the approximate form
\begin{equation}\label{ur-expected}
U_{ur}(\varphi)=\frac{m^2}{2}\varphi^{2}+\Delta U(\varphi)
\end{equation}
where $\Delta U(\varphi)$ should be a small quantity. 
\section{Minimal scalar field}\label{Sec3}
In this section, we reconstruct a general form of the scalar field potential in correspondence with the RRG. Firstly, we set up the basic relations between the scalar field and hydrodynamic variables and later we recover some basic properties of the RRG model through the equation of state in this scalar field context.
\subsection{Reconstructing the potential}
\begin{figure}[t!]
\centerline{\includegraphics[scale=0.66]{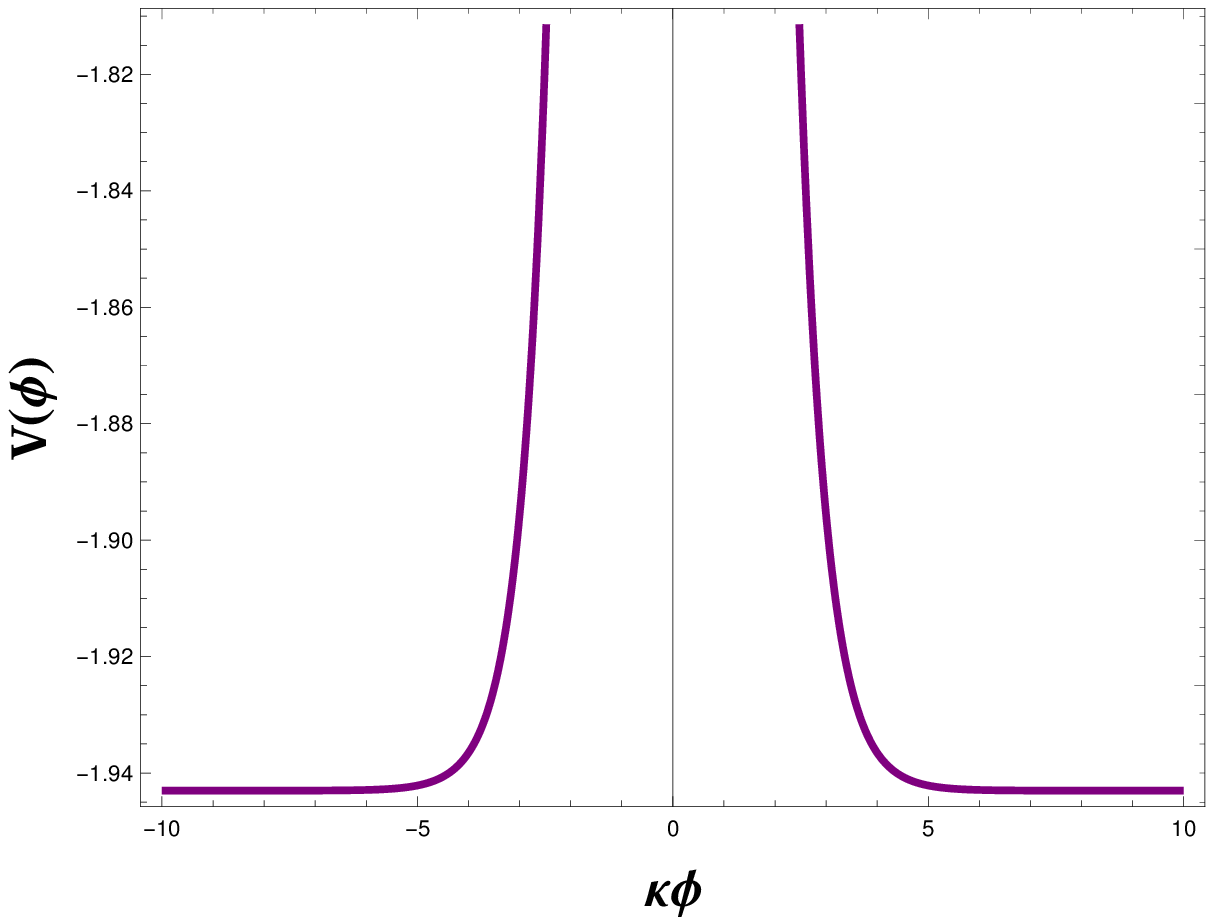}
\includegraphics[scale=0.7]{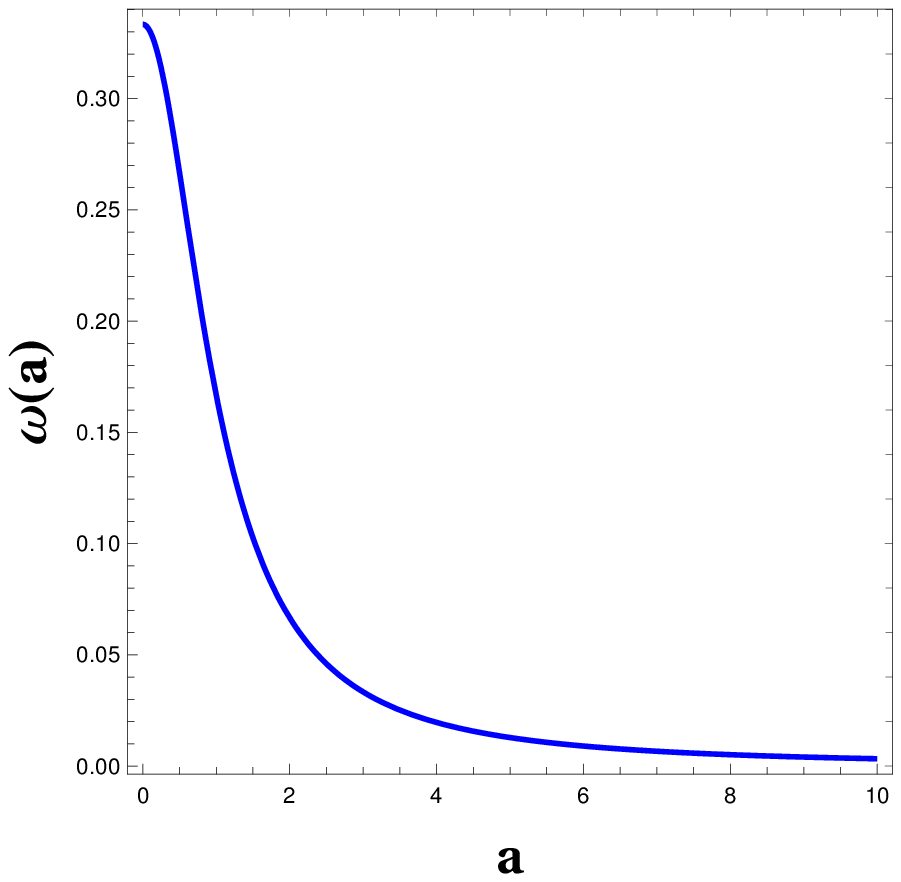}}
\begin{quotation}
\caption{\small \textit{Left panel:} The scalar potential $V(\phi)$ as a function of $\kappa\phi$. \textit{Right panel:} The equation of state $\omega(a)$ for the minimally coupled scalar field model for the RRG where is showed the interpolation between the radiation and matter domination phases.}\label{fig1}
\end{quotation}
\end{figure}

Consider the Lagrangian for a scalar field $\phi$  minimally coupled to gravity\footnote{We adopt here and what follows the next notations: $\phi$ is the scalar field minimally coupled to gravity and $\varphi$ is the scalar field conformally coupled to gravity.} \cite{quiros2019selected}
\begin{equation}\label{3.11}
S=\int{d^{4}x\sqrt{-g}\left[\frac{1}{2\kappa^2}R-\frac{1}{2}g^{\mu\nu}\partial_{\mu}\phi\partial_{\nu}\phi-V(\phi)\right]},
\end{equation}
with $\kappa^2=8\pi G=M_{pl}^{-2}$. The energy-momentum tensor for this scalar field is
\begin{equation}
    T^{\phi}_{\mu\nu}=\partial_{\mu}\phi\partial_{\nu}\phi-\left[\frac{1}{2}(\partial\phi)^{2}-V(\phi)\right]g_{\mu\nu}
\end{equation}
so using again the line element \eqref{metric}, the associated energy density and pressure are given by\footnote{When $\phi$ or $\varphi$ appear as indices will only make reference to minimal and conformal scalar field quantities, respectively, so they are not tensor indices.}
\begin{equation}\label{3.2}
\rho_\phi=\frac{1}{2}\dot\phi^2+V(\phi),\qquad\qquad p_{\phi}=\frac{1}{2}\dot\phi^2-V(\phi)
\end{equation}
and the conservation law $\nabla_\mu T^{\mu\nu}=0$ yields the Klein-Gordon equation
\begin{equation}
    \ddot{\phi}+3H\dot{\phi}+\frac{dV}{d\phi}=0.
\end{equation}
We want to construct a scalar field theory which describes the same cosmological evolution as the simple model for the RRG discussed in last section, following a similar procedure as was discussed, for example, in the case of Chaplygin gas \cite{kamenshchik2001alternative}. For this purpose, we need to make the equivalence
\begin{equation}\label{3.3}
\rho_R=\frac{1}{2}\dot\phi^2+V(\phi),\qquad\qquad p_{R}=\frac{1}{2}\dot\phi^2-V(\phi),
\end{equation}
such that
\begin{equation}\label{3.4}
\rho_R+p_{R}=\dot\phi^2,\qquad\qquad \rho_R-p_{R}=2 V(\phi).
\end{equation}
Using the first of these expressions for $\phi$, the Friedmann equations \eqref{friedmannrrg} and changing time derivatives by scale factor derivatives ($'\equiv d/da$), we would get
\begin{equation}\label{3.5}
\left(\frac{d\phi}{da}\right)^{2}=-\frac{1}{\kappa}\frac{1}{aH^2}\frac{dH^2}{da}=-\frac{1}{\kappa\,a\,\rho_R}\frac{d\rho_R}{da},
\end{equation}
and using now the conservation equation \eqref{RRGcosmeq1}
\begin{equation}\label{3.6}
\frac{d\phi}{da}=\frac{1}{\kappa\,a}\sqrt{4-s}.
\end{equation}
In order to simplify our calculations, let us to consider the redefiniton of the rest density in the form
\begin{equation}\label{3.7}
\rho_d=\rho_{1}a^{-3}\qquad\rightarrow\qquad\rho_d=\sqrt{3}\rho_1 a^{-3}
\end{equation}
such that equation \eqref{3.6} becomes
\begin{equation}\label{fund-rel}
\frac{d\phi}{da}=\frac{1}{2\kappa a}\left(\frac{a^2+4b^2}{a^2+b^2}\right)^{1/2},
\end{equation}
which is the fundamental and most general relation between minimal scalar field $\phi$ and the scale factor $a$, when a mapping between minimal scalar field and the RRG model is considered. Now, if we make the change of variable
\begin{equation}\label{3.10}
y^2=\frac{a^2+4b^2}{a^2+b^2},
\end{equation}
we obtain the next integral
\begin{equation}\label{3.11}
\phi(y)=\frac{1}{2\kappa}\int{\frac{6y^2dy}{(y^2-1)(y^2-4)}},
\end{equation}
whose solution is given by
\begin{equation}\label{3.12}
\phi(y)=\frac{1}{2\kappa}\ln{\left[\left(\frac{1+y}{1-y}\right)\left(\frac{2-y}{2+y}\right)^2\right]},
\end{equation}
or alternatively
\begin{equation}\label{3.13}
\cosh{2\kappa\phi}=-\frac{1}{2}\left(\frac{9y^{6}-15y^{4}+72y^{2}+16}{y^{6}-9y^{4}+24y^{2}-16}\right).
\end{equation}
After some simple algebra we can find a cubic equation for $y^2$, this is
\begin{equation}\label{3.16}
y^6-\lambda(\phi)y^4+\mu(\phi)y^2+\theta(\phi)=0,
\end{equation}
whose solution, which gives $y$ in terms of scalar field $\phi$, is given by
\begin{equation}\label{solutiony}
y^{2}(\phi)=\frac{1}{3}\left[\frac{2^{1/3}\chi(\phi)}{\iota(\phi)}+\frac{\iota(\phi)}{2^{1/3}}-\lambda(\phi)\right],
\end{equation}
where we have defined the functions
\begin{equation}\label{3.17}
\iota(\phi)=\psi(\phi)+\sqrt{\psi^{2}(\phi)-4\chi^{3}(\phi)},
\qquad\qquad
\chi(\phi)=\lambda^{2}(\phi)-3\mu(\phi),
\end{equation}
\begin{equation}
\psi(\phi)=-2\lambda^{3}(\phi)+9\lambda(\phi)\mu(\phi)-27\theta(\phi),
\end{equation}
with
\begin{equation}
\lambda(\phi)=\frac{3\left[1-12\cosh^{2}{\kappa\phi}\right]}{7+12\cosh^{2}{\kappa\phi}},
\qquad\qquad
\mu(\phi)=\frac{24\left[1+4\cosh^{2}{\kappa\phi}\right]}{7+12\cosh^{2}{\kappa\phi}},
\end{equation}
\begin{equation}
\theta(\phi)=\frac{16\left[3-4\cosh^{2}{\kappa\phi}\right]}{7+12\cosh^{2}{\kappa\phi}}.
\end{equation}
Therefore, we have found the solution of the scale factor $a$ as a function of the scalar field $\phi$ through the auxiliary variable $y$, namely
\begin{equation}\label{solutiona}
a^{2}(\phi)=b^{2}\frac{y^{2}(\phi)-4}{1-y^{2}(\phi)}.
\end{equation}
Additionally, solving the second equation in \eqref{3.4}, we can find the potential as a function of scale factor $a$, this is
\begin{equation}\label{phipot}
V(a)=\frac{\rho_1}{6a^4}\frac{5a^2+2b^2}{\sqrt{a^2+b^2}}.
\end{equation}
where substituting the solutions for $a(\phi)$ and $y(\phi)$ given by previous equation \eqref{solutiona} and \eqref{solutiony}, we can obtain the scalar potential $V(\phi)$, and therefore, a scalar field theory for the RRG in the case of minimal coupling. In figure \ref{fig1} (on the left) we plot this potential as a function of minimal scalar field $\kappa\phi$.
\subsection{UR-NR limits and equation of state}
In previous section we could obtain a general expression for the potential in this minimal scalar field context, but due the long and non-simple form of the function $y(\phi)$ in \eqref{solutiony}, the analytical treatment of this mapping can be complicated. However, we are interested in mapping our RRG model of WDM taking into account the behavior in the UR and NR limits, in order to compare our results, for instance, with the expected form of the conformal potential (see equation \eqref{ur-expected}). For this particular purpose, it is enough to consider the UR and NR version of fundamental relation \eqref{fund-rel}, such that
\begin{equation}\label{phipotur}
    UR:\quad\frac{d\phi}{da}=\frac{2}{\kappa a},\quad\Rightarrow\quad a_{ur}(\phi)=e^{\kappa\phi/2}\quad\Rightarrow\quad V_{ur}(\phi)=\frac{\rho_2}{3}e^{-2\kappa\phi}
\end{equation}
and
\begin{equation}\label{phipotnr}
    NR:\quad\frac{d\phi}{da}=\frac{1}{\kappa a},\quad\Rightarrow\quad a_{nr}(\phi)=e^{\kappa\phi}\quad\Rightarrow\quad V_{nr}(\phi)=\frac{5\rho_1}{6}e^{-3\kappa\phi}
\end{equation}
thus obtaining simpler expressions which can be employed in next section, when a conformal coupling for the scalar field is considered.\\
On the other hand, it would be interesting to check if the correspondent equation of state for this effective scalar field, it does satisfy the original properties of the equation of state for the simple RRG model, in these limits, as it was established in conditions \eqref{omegar}. In this scalar context, the equation of state is given by the usual expression  
\begin{equation}
\omega^{\phi}=\frac{p_\phi}{\rho_\phi}=\frac{\frac{1}{2}\dot\phi^2-V(\phi)}{\frac{1}{2}\dot\phi^2+V(\phi)}
\end{equation}
but using both Friedmann equations \eqref{friedmannrrg}, we can write
\begin{equation}\label{30}
\frac{1}{2}\dot{\phi}^{2}=-\frac{1}{\kappa^{2}}\frac{1}{aH^{2}}\frac{dH^2}{da}, \qquad\qquad V(\phi)=\frac{3H^2}{\kappa^2}+\frac{a}{2\kappa^2}\frac{dH^2}{da}
\end{equation}
and therefore we get
\begin{equation}
\omega^{\phi}=-1-\frac{a}{3H^2}\frac{dH^2}{da}=-1+\frac{1}{3}\left(\frac{3a^2+4b^2}{a^2+b^2}\right),
\end{equation}
where it is very clear now that, for the aforementioned limits, we simply obtain
\begin{equation}
\omega^{\phi}_{ur}=\frac{1}{3},\qquad\qquad \omega^{\phi}_{nr}=0,
\end{equation}
respectively. In the figure \ref{fig1} (on the right) we have plotted the equation of state parameter in this scalar context. Additionally, using equations in \eqref{30} we find that that the trace of energy-momentum tensor is
\begin{equation}\label{tphi}
T_{\phi}=4\,\dot{\phi}^2-V(\phi)=3\rho_{\phi}\left(\omega^{\phi}-1\right)
\end{equation}
which, in the UR limit, simply vanishes, such that, we can recover the properties of the original RRG model in this minimal scalar field formalism.
\section{Non-minimal scalar field}\label{Sec4}
In this section, we explore the possibility of completing our scalar field description of the RRG model but considering a more general action, where the conformal symmetry, as a connection between minimal and non-minimal cases, it receives special attention. We use simultaneous conformal transformation and reparametrization for constructing the non-minimal (conformal) potential from the minimal case.\\
\subsection{Conformal transformation and the general potential}
Let us consider now the general action for a scalar field theory in the form \cite{shapiro1995conformal}
\begin{equation}\label{4.1}
\bar{S}(\phi)=\int{d^4x\sqrt{-\bar{g}}\left[\bar{A}(\phi)\bar{g}^{\mu\nu}\partial_{\mu}\phi\partial_{\nu}\phi+\bar{B}(\phi)\bar{R}+\bar{C}(\phi)\right]}   
\end{equation}
where $\bar{A}$, $\bar{B}$ and $\bar{C}$ are arbitrary functions of $\phi$ and consider also the simultaneous conformal transformation with arbitrary field $\sigma=\sigma(\varphi)$ and scalar field reparametrization $\phi=\phi(\varphi)$, where
\begin{equation}\label{4.2}
\bar{g}^{\mu\nu}=g^{\mu\nu}e^{-2\sigma}, \qquad\qquad \sqrt{-\bar{g}}=\sqrt{-g}e^{4\sigma}
\end{equation}
\begin{equation}\label{4.3}
\bar{R}=e^{-2\sigma}\left[R-6\nabla^{2}\sigma-6(\nabla\sigma)^{2}\right]
\end{equation}
with 
\begin{equation}
  \nabla^{2}\sigma=g^{\mu\nu}\nabla_{\mu}\nabla_{\nu}\sigma, \qquad\qquad (\nabla\sigma)^{2}=g^{\mu\nu}\nabla_{\mu}\sigma\nabla_{\nu}\sigma.
\end{equation}
After applying these transformations and reparametrization, we could rewrite the action \eqref{4.1}, in the new variables as
\begin{multline}\label{4.4}
S(\varphi)=\int d^4x\sqrt{-g}\left\{ \left[e^{2\sigma}\bar{A}(\phi)\left(\frac{d\phi}{d\varphi}\right)^{2}+ \right.\right. 
\left.\left.+6e^{2\sigma}\Bigl[\bar{B}(\phi)\left(\frac{d\sigma}{d\varphi}\right)^2+\frac{d\bar{B}}{d\phi}
\frac{d\sigma}{d\varphi}\frac{d\phi}{d\varphi}\Bigr]\right]{g}^{\mu\nu}\partial_{\mu}\varphi\partial_{\nu}\varphi+\bar{B}(\phi)e^{2\sigma}{R}+e^{4\sigma}\bar{C}(\phi)\right\}
\end{multline}
such that we have the equivalence relations between both actions
\begin{equation}\label{rel1}
A(\varphi)=e^{2\sigma}\left\{\bar{A}(\phi)\left(\frac{d\phi}{d\varphi}\right)^{2}+6\left[\bar{B}(\phi)\left(\frac{d\sigma}{d\varphi}\right)^2+\frac{d\bar{B}}{d\phi}
\frac{d\sigma}{d\varphi}\frac{d\phi}{d\varphi}\right]\right\}
\end{equation}
\begin{equation}\label{rel2}
B(\varphi)=\bar{B}(\phi)e^{2\sigma},
\qquad\qquad
C(\varphi)=\bar{C}(\phi)e^{4\sigma},
\end{equation}
which can be used to find the explicit form of conformal field $\sigma(\varphi)$, the reparametrization $\phi=\phi(\varphi)$ and also the $C(\varphi)$, for two given theories related by the conformal transformation \cite{faraoni1998conformal}. In our case, let $\bar{S}(\phi)$ be the action for a minimally coupled scalar field, such that
\begin{equation}
\bar{A}(\phi)=-\frac{1}{2},\qquad\qquad \bar{B}(\phi)=\alpha,\qquad\qquad \bar{C}(\phi)=-V(\phi),
\end{equation}
where we left $\alpha$ as an arbitrary constant, which can be fixed later on. Additionally, let $S(\varphi)$ be the action for the conformally coupled scalar field model, so we demand that its corresponding functions take the explicit form
\begin{equation}
A(\varphi)=-\frac{1}{2},\qquad\qquad B(\varphi)=\frac{1}{2\kappa^2}-\frac{\xi}{2}\varphi^{2},\qquad\qquad C(\varphi)=-U(\varphi)
\end{equation}
where we will consider 
\begin{equation}
    U(\varphi)=U_{\xi}(\varphi){\Big\vert}_{\xi=1/6}
\end{equation}
in order to hold the conformal symmetry of the action $S(\varphi)$. Using the relation \eqref{rel2}, we can find the conformal field $\sigma$, which is given by
\begin{equation}\label{sigmaeq}
\sigma(\varphi)=\ln{\left[\frac{1}{2\alpha\kappa^2}\left(1-\frac{\kappa^{2}\varphi^{2}}{6}\right)\right]^{1/2}},
\end{equation}
and using \eqref{rel1}, we can solve for $\phi$ and to obtain its relation with $\varphi$, this is
\begin{equation}\label{rel4}
    \phi(\varphi)=\sqrt{3\alpha}\ln{\left(\frac{1+\frac{\kappa\varphi}{\sqrt{6}}}{1-\frac{\kappa\varphi}{\sqrt{6}}}\right)}=\frac{1}{\beta}\ln{\left(\frac{1+\frac{\kappa\varphi}{\sqrt{6}}}{1-\frac{\kappa\varphi}{\sqrt{6}}}\right)},
\end{equation}
where we have taken $\alpha=1/(3\beta^2)$ for convenience, letting $\beta$ as another arbitrary constant. Finally, using the last relation in \eqref{rel2}, we can find the function $U(\varphi)$, in the form
\begin{equation}
    U(\varphi)=\frac{9\beta^4}{4\kappa^4}V(\phi(\varphi))\left(1-\frac{\kappa^{2}\varphi^{2}}{6}\right)^2
\end{equation}
where the minimal potential $V(\phi(\varphi))$ was already determined in previous section according to equation \eqref{phipot}. 
\subsection{Explicit potential in the UR limit}
\begin{figure}[t]
\centerline{\includegraphics[scale=0.8]{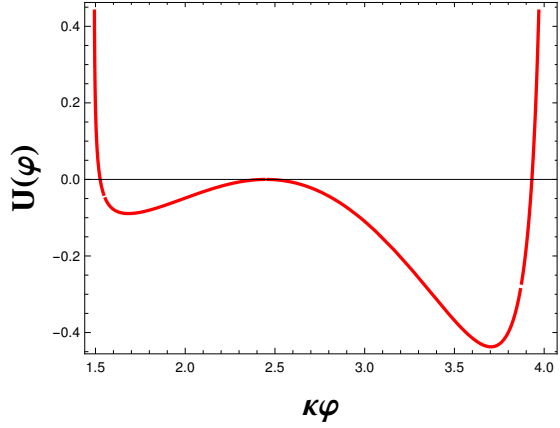}}
\begin{quotation}
\caption{\small The conformal potential $U(\varphi)$ as a function of $\varphi$.}\label{fig2}
\end{quotation}
\end{figure}

To obtain explicit formula for this potential, we need to substitute equation \eqref{rel4} in the solution for $a(\phi(\varphi))$ and take the result as an input into the expression for the potential \eqref{phipot}. In Figure \ref{fig2} we plot this potential as function of conformal field $\varphi$.\\
From the equation \eqref{sigmaeq}, it is clear that we have the condition
\begin{equation}
\frac{1}{2\alpha\kappa^2}\left(1-\frac{\kappa^{2}\varphi^{2}}{6}\right)>0,\qquad \Longrightarrow\qquad \varphi^2<\frac{3M_{p}^{2}}{4\pi}\approx\frac{M_{p}^{2}}{4}
\end{equation}
as expected for a classical scalar field.  
As mentioned at the end of section 2, let us to see what happen with this conformal potential once the UR limit is considered. Thus, using \eqref{limits} in \eqref{fund-rel}, we get again equation \eqref{phipotur}, namely
\begin{equation}
V_{ur}(\phi)=V(a_{ur}(\phi))=\frac{\rho_2}{3}\,e^{-2\kappa\phi}
\end{equation}
but using \eqref{rel4} on this UR version of the minimal potential, we get
\begin{equation}
    V_{ur}(\varphi)\approx\frac{\rho_2}{3}\left(\frac{1-\frac{\kappa\varphi}{\sqrt{6}}}{1+\frac{\kappa\varphi}{\sqrt{6}}}\right)^{2\kappa/\beta}.
\end{equation}
At this point it is possible to see that the only acceptable possibility for our arbitrary constant is $\beta=\kappa$, such that the conformal potential on this  UR limit takes the form
\begin{equation}\label{conf-pot-ur}
    U_{ur}(\varphi)=U_0(\varphi)+\Delta U(\varphi)
\end{equation}
where 
\begin{equation}
    U_0(\varphi)=\frac{3\rho_2}{4}\left((\kappa\varphi)^{2}-\frac{4(\kappa\varphi)}{\sqrt{6}}+1\right),\qquad\Delta U(\varphi)=\frac{3\rho_2}{4}\left(\frac{\kappa^{4}\varphi^{4}}{36}-\frac{2\kappa^{3}\varphi^3}{3\sqrt{6}}\right)\ll U_0(\varphi),
\end{equation}
as we can easily check, due both of these terms are proportional to $M_{p}^{-4}$ and $M_{p}^{-3}$, respectively, so $\Delta U(\varphi)$ is certainly small. Finally, it is possible to rewrite \eqref{conf-pot-ur} in terms of an auxiliary field defined as
\begin{equation}
\kappa\varphi=\sqrt{\chi^{2}-\frac{1}{3}}+\frac{2}{\sqrt{6}}
\end{equation}
such that
\begin{equation}
U_{ur}(\chi)=\frac{3\rho_2}{4}\chi^{2}+\Delta U(\chi)  
\end{equation}
also satisfying the condition
\begin{equation}
    \Delta U(\chi)\ll\frac{\rho_2}{\kappa^2}\chi^{2}
\end{equation}
exactly as was expected and in perfect agreement with the above equation \eqref{ur-expected} in section 2.\\
Additionally, due the the correspondent transformation of the energy-momentum tensor and its trace, which are given by
\begin{equation}
    T^{(\varphi)}_{\mu\nu}=e^{-2\sigma}T^{(\phi)}_{\mu\nu} \qquad\Rightarrow\qquad T^{(\varphi)}=e^{-4\sigma}T^{(\phi)}
\end{equation}
in the UR limit $T^{(\varphi)}$ also vanishes, as a consequence of the above result for $T^{(\phi)}$ as given by \eqref{tphi}. Therefore, we can reproduce the original UR limit for the RRG also in this non-minimal (conformal) scalar field description.
\section{Including running vacuum energy}\label{Sec5}
Let us explore the possibility of including a dark energy component in our scalar field model to unify the dark sector in a unique field. The simplest alternative is the CC, whose case only implies a constant contribution to the scalar field potential as given in \eqref{phipot} because its equation of state $p_{\Lambda}=-\rho_{\Lambda}$ does not modify the previous fundamental relation \eqref{fund-rel}. However, the situation can be very different if a dynamical form of dark energy is considered. In previous work, the author and collaborators have considered a cosmological model for an RCC and WDM, where it was studied the evolution of the model in both background and perturbative contexts and it was compared with some cosmic observables as the first acoustic peak and the matter power spectrum \cite{ruiz2020primordial,ruiz2020constraints}.\\
Let us consider the inclusion of this RCC for our scalar field model of WDM. To avoid a trivial running and to hold the WDM as a decoupled component, it is also needed to add a baryonic matter with an equation of state
\begin{equation}
    p_b=\omega_b \rho_b
\end{equation}
such that the Friedmann equations and the conservation law for this baryonic matter and RCC reads
\begin{equation}\label{frirrglambda}
    \rho_R+\rho_b+\rho_{\Lambda}=\frac{3}{\kappa^2}H^2,\qquad p_R+\omega_b \rho_b-\rho_{\Lambda}=-\frac{1}{\kappa^2}\left(3H^2+2\dot{H}\right)
\end{equation}
\begin{equation}
\dot{\rho}_b+3(1+\omega_b)\rho_b H=-\dot{\rho}_{\Lambda}
\end{equation}
where the running of the CC is given by the renormalization group equation, this is
\begin{equation}
\rho_\Lambda \,=\, \rho^0_\Lambda \,+\,
\frac{3\nu}{\kappa^2}\, \big( H^2 - H_0^2)=A+BH^2
\label{CCrun}
\end{equation}
with
\begin{equation}
    A= \rho^0_\Lambda-\frac{3\nu H_0^2}{\kappa^2},\qquad B=\frac{3\nu}{\kappa^2},
\end{equation}
and the WDM component evolves independently as given in \eqref{RRGcosmeq1} and \eqref{RRGcosmeq2}. In order to map this model to a minimally coupled scalar field, we can use these Friedmann equations in \eqref{frirrglambda} to write the relations \cite{basilakos2019scalar}
\begin{equation}
    \dot{\phi}^2=-\frac{2\dot{H}}{\kappa^2},\qquad V(\phi)=\frac{3H^2}{\kappa^2}+\frac{\dot{H}}{\kappa}
\end{equation}
or alternatively
\begin{equation}
    \phi'^{2}=-\frac{1}{\kappa^2 a}\frac{(H^2)'}{H^2},\qquad V(\phi)=\frac{3H^2}{\kappa^2}+\frac{a}{2\kappa^2}(H^2)'
\end{equation}
so we need to find an expression for the Hubble parameter. The general and analytical solution of this system of equations was presented in \cite{ruiz2020constraints}, where the Hubble parameter, in the case of null spatial curvature, takes the form
\begin{align}
\label{23}
H(a)^{2}
=&
H_{0}^{2}\Big[\Omega^{0}_{\Lambda}-\frac{\nu}{1-\nu}\left(\Omega^{0}_{b}+\Omega^{0}_{dm}\right)+\frac{a^{-\zeta}}{1-\nu}\left(\Omega^{0}_{b}+\nu\Omega^{0}_{dm}\right)+\frac{\Omega^{0}_{dm}a^{-3}}{\sqrt{1+b^2}}\sqrt{1+\frac{b^2}{a^2}}\\&+{}_2 F_1(f_1, f_2; f_3; -b^2)Ca^{-\zeta}-{}_2 F_1(f_1, f_2; f_3; Z)Da^{-3}\Big]
\end{align}
where ${}_2 F_1(f_1, f_2; f_3; Z)$ is the hypergeometric function
defined as
\begin{equation}\label{19}
{}_2 F_1(f_1, f_2; f_3; Z) = \sum_{k=0}^\infty \frac{(f_1)_k (f_2)_k}{(f_3)_k}
\,\frac{Z^k}{k!},
\end{equation}
$(\alpha)_k\,$ is the Pochhammer symbol and we have defined the constants
\begin{equation}
\label{20}
f_1 = -\frac12,
\qquad
f_2 = \frac{3-\zeta}{2},
\qquad
f_3 = \frac{5-\zeta}{2}
\qquad
Z = -b^2/a^2.
\end{equation}
\begin{equation}
    C=\frac{\nu\zeta\Omega^{0}_{dm}}{(1-\nu)(3-\zeta) },\qquad D=\frac{C}{\sqrt{1+b^2}},
\qquad    \zeta=3(1+\omega_b)(1-\nu).
\end{equation}
The complete solution for the scalar field mapping in this case, using the general expression for $H(a)$ implies two technical steps : integrating the terms evolving the hypergeometric function and inverting the resulting expression to get $a=a(\phi)$. We let the details and discussion about this full treatment for a future work.  For now, and in order to compare the results with previous section \ref{Sec3}, we can take the UR limit in $H(a)$, thus obtaining simpler expressions and mapping our model to a scalar field for this particular case. In this limit we can also take $|\nu|\approx 0$, such that this expression simply reduces to
\begin{equation}
    H_{ur}(a)=H_{0}^{2}\left(\Omega^{0}_{\Lambda}+\Omega^{0}_{m}a^{-4}\right)
\end{equation}
where $\Omega^{0}_{m}=\Omega^{0}_{b}+\Omega^{0}_{dm}$. Therefore, the first equation of the scalar field mapping yields
\begin{equation}
    \frac{d\phi}{da}=\frac{2/\kappa a}{\sqrt{\left(\frac{\Omega^{0}_{\Lambda}}{\Omega^{0}_{m}}\right)a^{4}+1}}
\end{equation}
and the potential takes the form
\begin{equation}
    V_{ur}(a)=\frac{H_{0}^{2}}{2\kappa^2}\left(3\Omega^{0}_{\Lambda}+\Omega^{0}_{m}a^{-4}\right)
\end{equation}
where, in order to compare with our previous result, we have to take the limit $\Omega^{0}_{\Lambda}\rightarrow 0$ and $\Omega^{0}_{b}\rightarrow 0$, thus obtaining exactly the expressions \eqref{phipot} and \eqref{phipotur}. 
\section{Conclusions}
\label{Sec6}
We have studied here the scalar field theory for a warm dark matter (WDM) component which is modeled using the simple reduced relativistic gas approximation (RRG). As a general result, we have found a mapping between the RRG model and a scalar field minimally coupled to gravity. Additionally, we have reconstructed the non-minimal action and potential from this minimal case. For the minimal coupling between gravity and the scalar field $\phi$, we found the correspondent scalar potential and analyze the behavior of its analogous equation of state and the trace of the energy-momentum tensor, where the properties of the original RRG model were verified in terms of this minimal scalar field.\\
In the context of non-minimal coupling to gravity, we have constructed the conformal potential from the minimal case using the properties of the conformal transformation of the metric tensor, where, as expected, we could verify the consistency of this general scalar field description in the UR limit and for the case of conformal invariance of the action $S(\varphi)$ ($\xi=1/6$). Our results suggest that the RRG approximation can be successfully replaced by the scalar models, in terms of simple minimal and conformal scalar fields.\\
We also explore the possibility of making a similar map including a dynamical dark energy component in the form of a running cosmological constant (RCC). It was possible to verify in the UR case and for the limit of null baryon and vacuum energy densities, the correspondence with the aforementioned pure WDM case. The complete mapping for this model, considering the most general expression for the Hubble parameter $H(a)$ and taking into account a non-zero running parameter $\nu$ will be considered in future work. The scalar field mapping of this RCC could be especially interesting because its original formulation does not admit an explicitly covariant description and therefore represents a difficult challenge for the analysis of cosmic perturbations, especially CMB and gravitational waves. 

\section*{Acknowledgements}
\noindent J. A. A. Ruiz thanks to CAPES for supporting his Ph.D. project. The author is also very grateful to I. L. Shapiro, L. Giani and E. Frion for useful comments and critical observations.
\bibliographystyle{unsrt}
\bibliography{Refs}
\end{document}